# Comparative study of terbium tellurides Tb$_2$Te$_5$ and TbTe$_3$

I. Shamova,[1] J.-Y. Sun,[2] X.-Y. Chang,[2] V. Popova,[1] D. Chareev,[3,4,5] L. Shvanskaya,[1,6] D. Ksenofontov,[6] A. Vorobyova,[6,7] K. Lyssenko,[6] A. Demidov,[8] H.-Y. Yeh,[2] W.-Y. Tzeng,[9] J.-Y. Lin,[10] C-W. Luo,[2,10,11] P. Monceau,[12] E. Pachaud,[12] E. Lorenzo,[12] A. Sinchenko,[13,14] A. Vasiliev,[1,6] O. Volkova[6]

[1]National University of Science and Technology MISIS, Moscow 119049, Russia
[2]Department of Electrophysics, National Yang Ming Chiao Tung University, Hsinchu 30010, Taiwan
[3]Korzhinskii Institute of Experimental Mineralogy, RAS, Chernogolovka 142432, Russia
[4]Dubna State University, Dubna 141980, Russia
[5]Ural Federal University, Ekaterinburg 620002, Russia
[6]Lomonosov Moscow State University, Moscow 119991, Russia
[7]Higher School of Economics, Moscow 101000, Russia
[8]Bryansk State Technical University, Bryansk 241035, Russia
[9]Department of Electronic Engineering, National Formosa University, Yunlin 632, Taiwan
[10]Institute of Physics, National Yang Ming Chiao Tung University, Hsinchu 300093, Taiwan
[11]National Synchrotron Radiation Research Center, Hsinchu 30092, Taiwan
[12]University Grenoble Alpes, Institute Neel, F-38042 Grenoble, France and CNRS, Institute Neel, F-38042 Grenoble, France
[13]Laboratoire de Physique des Solides, Université Paris-Saclay, CNRS, 91405 Orsay Cedex, France
[14]Kotelnikov Institute of Radioengineering and Electronics of RAS, 125009 Moscow, Russia

Two terbium tellurides, TbTe$_3$ and Tb$_2$Te$_5$, were studied by means of thermodynamics, ultrafast pump-probe spectroscopy and torque magnetometry. While the crystal structure and some physical properties of TbTe$_3$ were established previously, the crystal structure of Tb$_2$Te$_5$ is reported based on the low-temperature (100 K) single crystal X-ray diffraction experiment. Tb$_2$Te$_5$ crystallizes in orthorhombic *Cmcm* space group with $a$ = 4.3009(17), $b$ = 43.107(18), $c$ = 4.3016(17)Å, $V$ = 797.5(6) Å$^3$, $Z$ = 4. In contrast to TbTe$_3$, which experiences three successive magnetic phase transitions, Tb$_2$Te$_5$ orders antiferromagnetically in two steps at $T_{N1}$ = 9.0 K and $T_{N2}$ = 6.8 K, both readily suppressed by an external magnetic field. The third transition in TbTe$_3$ is due to the interaction of the magnetic subsystem with the charge density waves. The interaction of magnetic and electronic subsystems in Tb$_2$Te$_5$ has been revealed by the pump probe. Torque measurements of TbTe$_3$ show that the magnetic moments of Tb are oriented predominantly in the *ac* plane at high temperatures and switch to the *b* axis at low temperatures. In Tb$_2$Te$_5$, the magnetic moments of Tb are oriented predominantly in the *ac* plane at low temperatures.

2-dimensional systems; Antiferromagnets; Charge density waves; Torque magnetometry

**Introduction**

Due to their peculiar physical properties and high heat resistance, the rare-earth (RE) compounds are the important functional materials widely used in aerospace engineering, computer hard drives, powerful solid-state lasers and infrared sources [1]. Especially well-suited to modern technologies are the layered systems, which could be easily integrated into the planar elements of electronic devices. Among these systems are those of RE$_2$Te$_5$ and RETe$_3$ [2]. The members of these families are stabilized structurally by the van der Waals forces and exhibit an amazing variety of quantum cooperative effects, i.e., charge density waves (CDW) [3], exotic magnetism [4] and superconductivity under pressure [5]. Interaction of these orders results sometimes in very complicated magnetic phase diagrams as was observed recently in TbTe$_3$ [6].

This van der Waals system consists of the buckled layer of TbTe separated by the double planar net of Te perpendicular to the *b* axis of its weakly orthorhombic structure. Below room temperature, the tellurium layers are modulated by the charge density waves [7]. Antiferromagnetic phase transitions in TbTe$_3$ at $T_{N1}$ = 6.7 and $T_{N2}$ = 5.7 K are accompanied by formation of commensurate magnetic structures with wave vectors q$_{m1}$ = (0.5, 0.5, 0) and q$_{m2}$ = (0, 0, 0.5). The low temperature antiferromagnetic structure below $T_{N3}$ =5.4 K is incommensurate and its wave vectors can be presented as a sum of q$_{m1}$ and q$_{m2}$ with charge density wave vectors $q_c$ = (0, 0, 0.296) and $q_a$ = (0.32, 0, 0) observed below 330 K and 40 K, correspondingly [7, 8].

Compared to well-studied terbium tri-telluride, its sister system Tb$_2$Te$_5$ is virtually uninvestigated, there has been no information on its crystal structure and properties. To fill this gap, we undertook a combined experimental ant theoretical study of physical properties of Tb$_2$Te$_5$ focusing on magnetism.

**Crystal growth and X-Ray diffraction study**

The crystals of Tb$_2$Te$_5$ were grown by evaporating of liquid tellurium from a tellurium-terbium liquid in a boomerang-shaped quartz glass ampoule as described previously [9]. The ratio of Tb:Te=1:2.5 was confirmed by energy-dispersive X-ray spectroscopy.

To structurally characterize Tb$_2$Te$_5$ we have performed single crystal XRD study. Single crystal was investigated on a Bruker D8 QUEST single-crystal X-ray diffractometer equipped with PHOTONII detector, charge-integrating pixel array detector (CPAD), laterally graded multilayer (Goebel) mirror and microfocus Mo-target X-ray tube ($\lambda$ = 0.71073 Å). It should be noted that despite of numerous sample only one suitable single crystal was found for the experiment, although the diffraction pattern in the reciprocal space for this sample can be interpreted as growth twin or the consequence of turbostratic disorder. The latter almost exclude the possibility to find smaller and less defect crystal because the cutting of the crystal inevitably cause increase of turbostratic disorder. According to XRD at 100 K Tb$_2$Te$_5$ crystallizes in orthorhombic *Cmcm* space group with $a$ = 4.3009(17), $b$ = 43.107(18), $c$ = 4.3016(17) Å, V= 797.5(6) Å$^3$, Z = 4, $\mu$(MoK$\alpha$) = 354.1 cm$^{-1}$, d$_{calc}$ = 7.961 gcm$^{-3}$. For Tb$_2$Te$_5$ a total of 1591 (2$\theta_{max}$ = 50°, Rint = 0.0692) reflections were collected and 433 independent reflections were used for the structure solution and refinement, which converged to R$_1$ = 0.0894 (for 374 observed reflections and 30 parameters), $w$R$_2$ = 0.20931, GOF = 1.093. Absorption correction was applied using multiscan technique as implemented in SADABS [10]. Crystal structure solution and refinement were performed using SHELX-2018 package [11]. Crystal data and details of data collection and refinement are presented in Table S1 [12]. Atomic positions were located using dual methods and refined using least-square in anisotropic approximations. The observed turbostratic disorder cause some systematic mistakes in intensities and unexpected variation of anisotropic displacement parameters that cannot be corrected by the refinement of self occupancy factors. Despite the above problems, the experiment allows us to unambiguously prove the composition and crystal structure of Tb$_2$Te$_5$. In Table S2, S3 we report the final results of the atom positions with equivalent isotropic displacement parameters and characteristic distances, respectively.

According to X-Ray diffraction analysis, the crystal structure of Tb$_2$Te$_5$ belongs to the Nd$_2$Te$_5$ structural type [13]. In Tb$_2$Te$_5$ crystal structure Tb atoms occupy two non-equivalent positions: Tb1, Tb2. Each of these positions is coordinated by nine Te atoms, forming unicapped tetragonal antiprism. The Tb-Te bond lengths lie in the intervals 3.181(2) – 3.276(5) and 3.151(6) – 3.264(8) Å (Table S2) for Tb1 and Tb2 - centered polyhedra, respectively. Tb1O$_9$ and Tb2O$_9$ share edges to form four-layer packages that alternate along the *b* axis (Fig. 1, left).

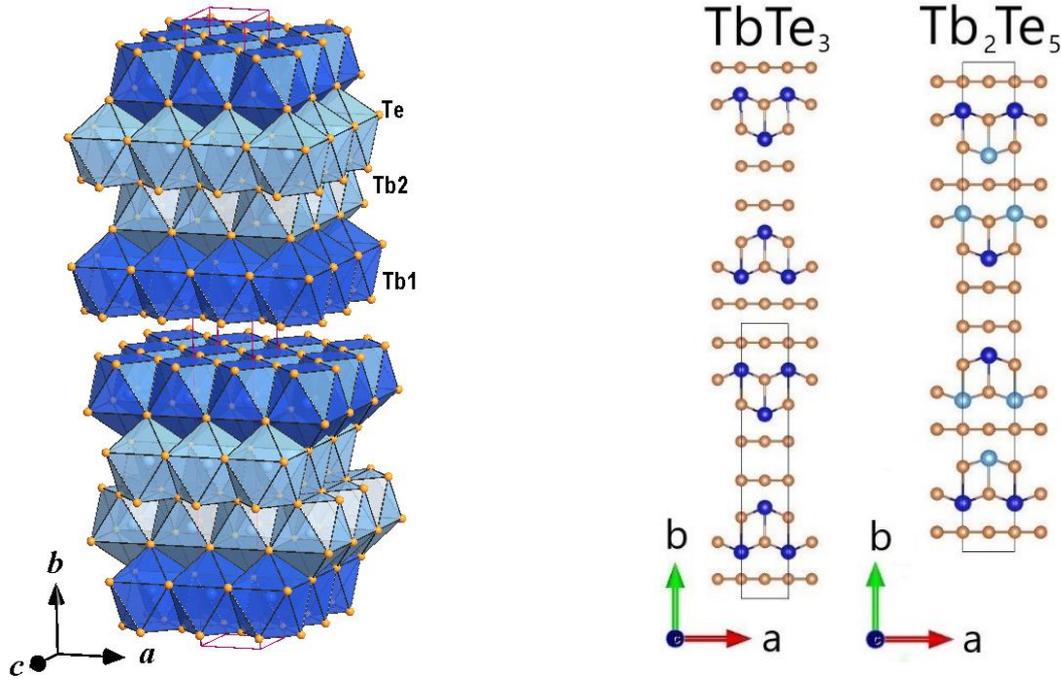

Fig. 1. Left panel: crystal structure of $Tb_2Te_5$ in polyhedra representation. Right panel: in Ball – and – stick representation of quasi-layerered crystal structures of $TbTe_3$ and $Tb_2Te_5$. Large spheres of different colors represent two crystallographically inequivalent sites of Tb in $Tb_2Te_5$.

X-ray powder diffraction study was provided on a STOE-STADI MP (Germany) diffractometer with a curved Ge (111) monochromator using CoKα radiation (λ = 0.178897 Å). The recorded X-ray powder diffraction pattern is shown in Fig. S1. All profile and lattice parameters were refined by Le Bail method [14] using the JANA2006 program [15]. Most of the peaks are indexed to *Cmcm* space group. Two additional impurity peaks, seen at 34° and 45°, can be attributed to the Te phase (ICSD-65692). The content of $Tb_2Te_5$ and Te has been estimated as 84(1) and 16(1)%, respectively. The refined values of their cell parameters are equal to $a$ = 4.3184(15), $b$ = 43.303(15) and $c$ = 4.3035(15) Å, α = β = γ = 90°, sp.gr. *Cmcm* ($Tb_2Te_5$); $a$ = 4.457(9), $c$ = 5.910(18) Å, α = β = 90°, γ = 120°, sp.gr. $P3_12_1$ (Te).

The crystal structures of $TbTe_3$ and $Tb_2Te_5$ are shown in the right panel of Fig. 1 in ball-and-stick representation. In variance with $TbTe_3$ structure, in the buckled TbTe layers are separated alternatively by either single or double square Te net. The $TbTe_3$ crystals were prepared by the method described in Ref. [6].

**Thermodynamics**

Thermodynamic measurements, i.e. magnetization and specific heat, were done on square thin plates of $Tb_2Te_5$ using various options of "Quantum Design" MPMS 7T and PPMS 9T devices. Thermodynamic properties of $TbTe_3$ single crystal and its magnetic phase diagram were established in Ref. 6. The temperature dependences of magnetic susceptibility of $Tb_2Te_5$ crystal in a magnetic field oriented either parallel or perpendicular to the *ac* plane are shown in the left panel of Fig. 2. At high temperatures, these curves follow the Curie – Weiss law with inclusion of the temperature-independent term, i.e., $\chi = \chi_0 + C/(T-\Theta)$ seen as high temperature linear type of behavior of $1/(\chi - \chi_0)$ curves in the inset to left panel of Fig. 2. The parameters of the fit taken in the temperature range 150 -300 K are $\chi_{0\parallel}$= -4.6·10$^{-3}$ emu/mol, $C_\parallel$ = 24.4 emuK/mol, $\Theta_\parallel$ = -50 K and $\chi_{0\perp}$= -2.6·10$^{-3}$ emu/mol, $C_\perp$ = 23.9 emuK/mol, $\Theta_\perp$ = 11 K. The effective magnetic moments, $\mu_{eff} = \sqrt{8C}$, equals to $\mu_{eff\parallel}$ = 14.0 $\mu_B$/f.u. and $\mu_{eff\perp}$ = 13.8$\mu_B$/f.u., which is close to the calculated value of effective magnetic moment of two $Tb^{3+}$ ions per formula unit $\mu_{eff}^{calc}$ = 13.8$\mu_B$/f.u. The field dependence of powder $Tb_2Te_5$ compound demonstrate a metamagnetic transition at $B_M$=2.9

T as is shown in the middle panel of Fig. 2. For magnetic field oriented along the *ac*-plane and *b* - axis the field of the metamagnetic transitions constituted 3.7 T and 2.5, correspondingly. The value of magnetization for $B||ac$ and $B||b$ approaches 7μ$_B$/f.u. and 11.5 μ$_B$/f.u. It is evident that at 2 K, in low fields, inset to the middle panel of Fig. 2, magnetization in the *ac*-plane prevails that along the *b* – axis, i.e. $M_{ac} > M_b$.

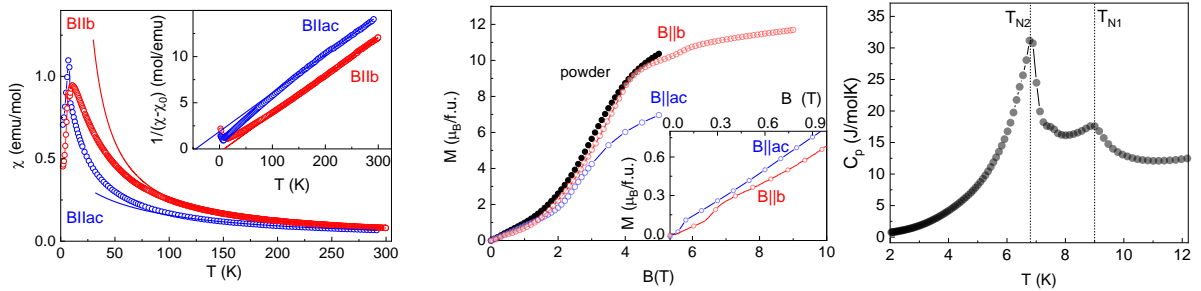

Fig. 2. The left panel represents temperature dependences of magnetic susceptibility and reversal magnetic susceptibility in the inset of crystalline Tb$_2$Te$_5$ taken in field – cooled regim at $B = 0.1$ T. The middle panel represents field dependences of magnetization of powder (closed circles) and oriented crystal (open circles) of Tb$_2$Te$_5$ at 2 K. The inset enlarges the low-field region. The right panel represents temperature dependence of specific heat of Tb$_2$Te$_5$ crystal.

The temperature dependence of the specific heat in Tb$_2$Te$_5$ evidences two anomalies at $T_{N1}$ = 9.0 K and $T_{N2}$ = 6.8 K, as shown in the right panel of Fig. 2. Similar to behavior of numerous representatives of RETb$_3$ family, these two transitions associate with magnetic long-range order at $T_{N1}$ and spin-reorientation at $T_{N2}$. Notably, three phase transitions in TbTe$_3$ [6] and four phase transitions in Ce$_2$Te$_5$ [16,17] were seen due to the interaction of magnetic order with charge density waves. The sets of temperature and field dependences of magnetization (see Figs. S2 – S3 in The Supporting Information) were used to construct magnetic phase diagrams of Tb$_2$Te$_5$ in a magnetic field oriented either perpendicular or within the *ac* plane, as shown in Fig. 3.

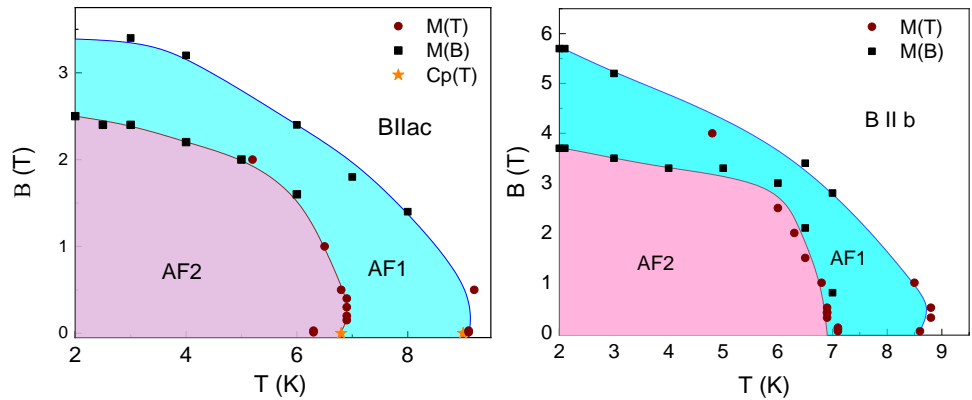

Fig. 3. Magnetic phase diagrams of Tb$_2$Te$_5$ crystal with magnetic field oriented $B||ac$ (left panel) and $B||b$ (right panel).

**Ultrafast pump-probe spectroscopy**

Although thermodynamic measurements did not reveal an interaction between magnetic order and the charge-density wave (CDW), ultrafast pump-probe spectroscopy reveals a coupling between electrons in the tellurium layers and the localized terbium magnetic moments within the buckled TbTe layers of Tb$_2$Te$_5$, as detailed below. The experiments employed a Ti:sapphire laser system (Femtosource scientific XL300, Femtolaser) generating 100 fs pulses centered at 800 nm with a 5.2 MHz repetition rate. The laser output was split into pump and probe beams. The pump beam wavelength was frequency-doubled to 400 nm using a β-BaB$_2$O$_4$ (BBO) crystal after passing through an acousto-optic modulator (AOM) for intensity modulation and lock-in detection [18]. A mechanical delay stage controlled the time delay ($\Delta t$) between pump and probe pulses. To measure the transient reflectivity change ($\Delta R/R$), the 400 nm pump and 800 nm probe pulses were cross-polarized to minimize coherent artifacts around $\Delta t = 0$ [19]. Typical pump and probe fluences were 41.6 µJ/cm$^2$ and 3.5 µJ/cm$^2$, respectively. Temperature-dependent measurements were conducted with the sample mounted in a cryostat.

The static reflectance spectrum of Tb$_2$Te$_5$ is shown in Fig. 4(c). The relatively low reflectance at 400 nm (3.1 eV) facilitates strong absorption via interband transitions, while the high reflectance at 800 nm (1.55 eV) allows probing of the ultrafast dynamics, including the shift and weight changes of the absorption spectra due to the variation in temperature and the photoexcited carrier density. Figure 4(a) presents typical $\Delta R/R$ transients at various temperatures. Following photoexcitation, $\Delta R/R$ exhibits a rapid rise due to the generation of photoexcited carriers, concurrent with ultrafast demagnetization. The subsequent relaxation involves multiple processes. An initial fast decay (positive $\Delta R/R$, red shading in Fig. 4(a). For the details, please see Fig. S4 and its relative explanations), occurring within picoseconds, is attributed to carrier relaxation primarily through electron-phonon coupling. Subsequently, a slower relaxation process emerges (timescale ~100 ps; negative $\Delta R/R$, blue shading in Fig. 4(a). For the details, please see Fig. S4 and its relative explanations), which is associated with the recovery of magnetic order [18]. Closer examination of the early-time dynamics [Fig. 4(d)] reveals a distinct oscillatory component superimposed on the fast electronic relaxation. The frequency of this oscillation is determined to be 3.77 THz via Fourier analysis [inset, Fig. 4(d)]. Such oscillations are recognized signatures of the CDW amplitude mode, previously observed in related tellurides such as $R$Te$_3$ ($R$=La, Ho, Dy, Tb) and other CDW materials like K$_{0.3}$MoO$_3$ and CuTe [20]. The observed frequency is higher than the typical range (1.5–2.5 THz) for $R$Te$_3$ amplitude modes, likely due to structural differences between $R$Te$_3$ and Tb$_2$Te$_5$. The presence of this 3.77 THz oscillation within the positive $\Delta R/R$ regime (< 5 ps) provides strong evidence for a CDW phase in Tb$_2$Te$_5$.

The zero-crossing ($\Delta R/R = 0$ with white color in Fig. 4(a)) delay time, $t_m$, shows the significant changes around the magnetic phase transition temperatures $T_{N1} = 9.0$ K. The quantitative analyses are presented in Fig. 4(b) and its inset. The $t_m$ almost linearly decreases as decreasing temperature until 9 K. Surprisingly, the $t_m$ suddenly drops below 9 K due to the coupling between the electron subsystem and magnetic subsystem, which strongly indicates the appearance of an antiferromagnetic (AFM) phase, associated with the negative $\Delta R/R$ and slow relaxation process with timescale of ~100 ps, at $T_{N1}$. The $t_m$ stabilizes at about 4.6 ps below $T_{N2}$. These results show the close correlation between the magnetic subsystem, i.e., the AFM phases, and the electron subsystem in Tb$_2$Te$_5$. A smaller $t_m$ below $T_{N1}$ implies a slower initial disruption (i.e., the rising time $t_{dp}$ of the negative component in Fig. 4(b)), potentially reflecting the robustness of the established long-range AFM order at lower temperatures, which requires more energy or time to perturb significantly after photoexcitation. Conversely, the gradual decrease of $t_m$ upon cooling above $T_{N1}$ might suggest the influence of growing short-range AFM correlations, which can also be revealed by the temperature-dependent negative component of $\Delta R/R$ above $T_{N1}$ in Fig. S5 (see Supplementary Materials). This interpretation aligns with similar pump-probe studies on magnetic ordering dynamics in other materials, including multiferroics [21].

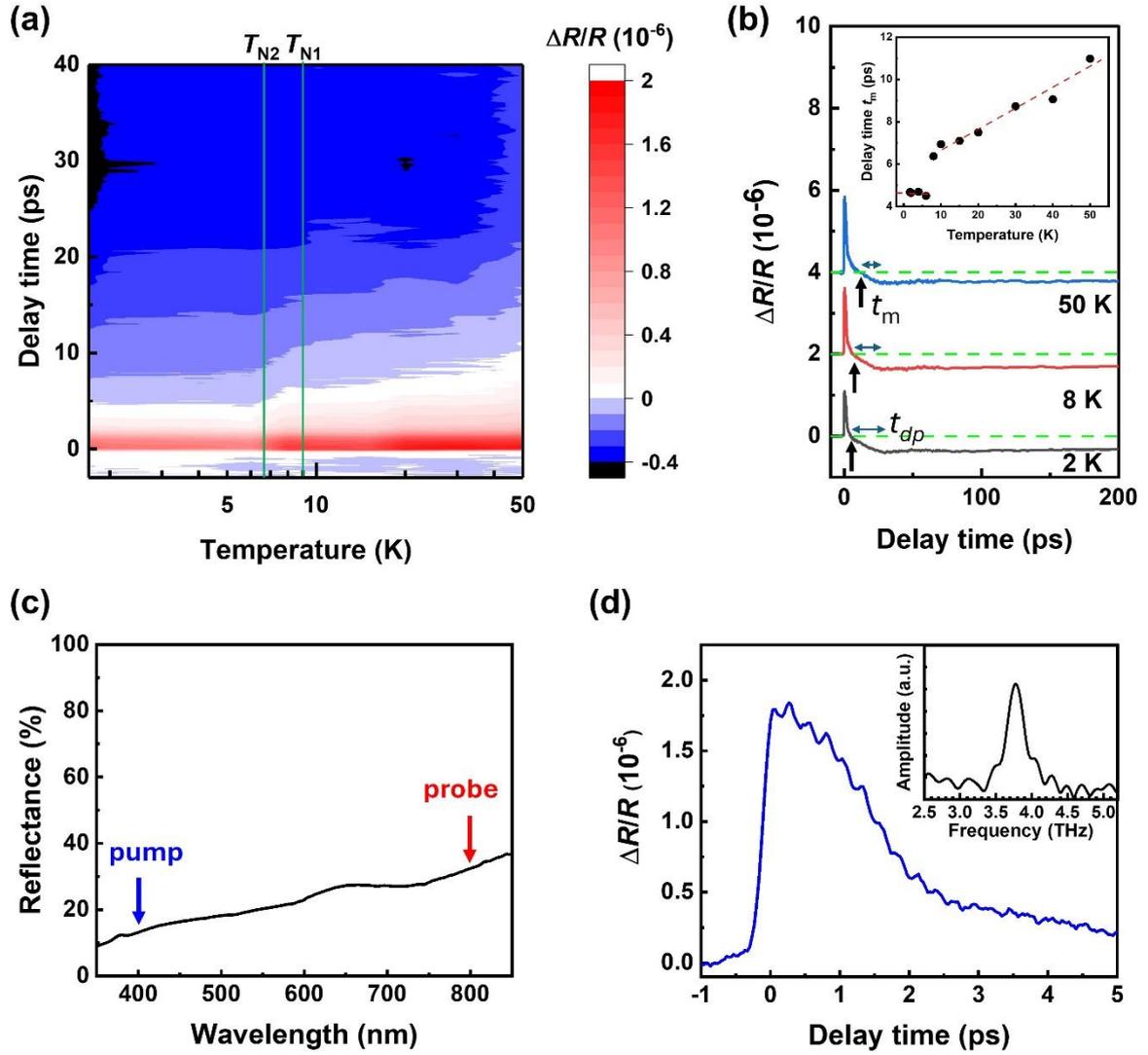

Fig. 4. (a) Transient reflectivity change ($\Delta R/R$) spectra of a $Tb_2Te_5$ crystal at various temperatures. (b) $\Delta R/R$ spectra in (a) with longer delay time. Inset: the zero-crossing ($\Delta R/R = 0$) delay time $t_m$, marked by an arrow in (b), as a function of temperatures. Dashed lines guide the eye above 8 K and below 6 K. (c) Reflectance spectrum of a $Tb_2Te_5$ crystal at room temperature. (d) Zoomed plot (around zero delay time) of the $\Delta R/R$ spectrum of a $Tb_2Te_5$ crystal at 50K in (a). Inset: Fourier transformation spectrum of (d).

**Torque magnetometry**

Measurements of torque were done on rectangular plate crystals by "Quantum Design" Torque magnetometer chip combined with a rotator installed in PPMS 9T device. $TbTe_3$ crystals possess facets perpendicular to *a* and *c*-axes. As shown in Fig. 5, an external magnetic field has been rotated within either *ab* or *cb* – planes. For $Tb_2Te_5$ we found the misorientation of crystallites within the *ac* plane. The φ = 0° and 90° correspond to B ∥ *b* and B ∥ *ac*.

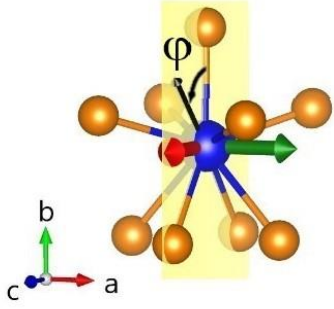

Fig. 5. TbTe$_9$ polyhedron constituting TbTe$_3$ and Tb$_2$Te$_5$ structures. Red and green arrows indicate easy directions of magnetic moment of Tb$^{3+}$ in the *ac*-plane. The magnetic field, shown by black arrow, rotates either in the *bc* plane, shown in yellow, or in the *ab* plane. The φ angle was measured from the *b* – axis.

The azimutal torque dependencies for both TbTe$_3$ and Tb$_2$Te$_5$ are summarized in Fig. 6. The torque τ of purely axial systems without anisotropy in the *ac* plane can be described by the equation [22]

$$\tau = M_b B_{ac} - M_{ac} B_b = B(M_b \sin\varphi - M_{ac}\cos\varphi), \qquad (1)$$

where $M_b$ and $M_{ac}$ are the perpendicular and parallel components of magnetization within the rotation plane and φ is the angle of rotation. The zero values of τ at 90° and 180° indicate that magnetic field *B* is either parallell or perpendicular to the easy direction in both systems.

As shown in the left panel of Fig.6, the negative sign of τ at φ=45° in TbTe$_3$ corresponds to $M_b - M_{ac} < 0$, i.e. to the *ac* – easy – plane. The torque data are in good correspondence with the magnetization curves taken both above and below $T_{N3}$. Magnetization curves of TbTe$_3$ in weak fields reveal almost temperature independent behavior of $M_b(B)$ dependences and strong transformation of $M_c(B)$ and $M_a(B)$ T< $T_{N3}$, as is shown in Fig. S6 in Supplementary Materials. At T >$T_{N3}$, the magnetization in the *ac* – plane exceeds that along the *b* – axis, ($M_c, M_a > M_b$), leading to the negative torque at 45°, τ(45°)~$M_b$-$M_{ac}$ < 0. At T ~ $T_{N3}$, the $M_a(B)$, $M_c(B)$ and $M_b(B)$ curves are close to each other which results in complex behavior of the τ(ϕ) dependencies at lower temperatures, shown in left panel of Fig. 6. At certain φ either $M_{ac} > M_b$ or $M_{ac} < M_b$. At lowest temperatures, $M_a(B), M_c(B) < M_b$ and τ (45°) >0. Theoretical calculation of torque of TbTe$_3$, shown by solid lines in Fig. 6, is given in next section.

The azimuthal dependences of torque of Tb$_2$Te$_5$ demonstrate zero values at 90° and 180° at all temperatures. At $T < 14$ K, τ(45°) is negative, i.e. $M_b < M_{ac}$. Thus, easy *ac* – plane anisotropy can be assumed for Tb$_2$Te$_5$ similar to the relative TbTe$_3$. This behavior is preserved in the magnetically ordered state as is shown in the right panel of Fig. 6.

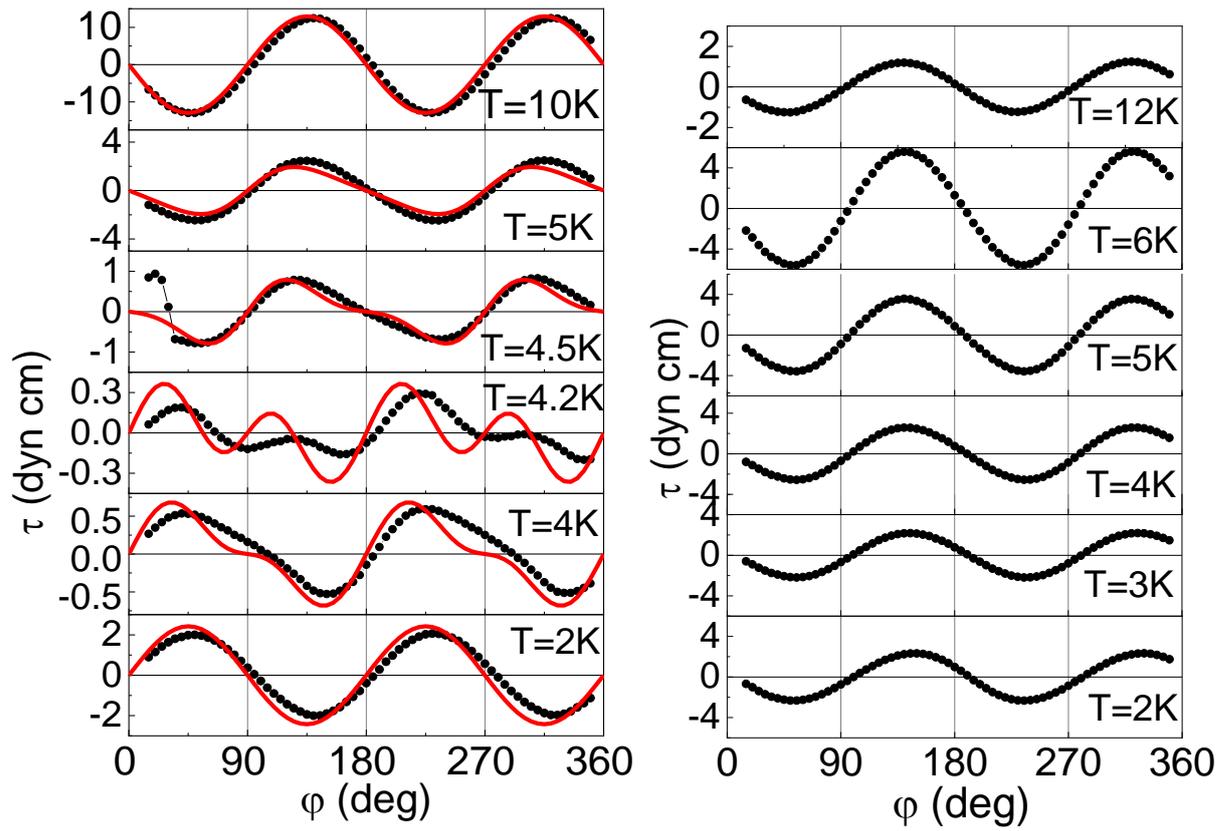

Fig. 6. Torque signals of TbTe$_3$ (left) and Tb$_2$Te$_5$ (right) at various temperatures at $B = 0.5$ T. Solid line is a fit.

**Calculations of magnetization and torque of TbTe₃**

In present calculations we used the approach developed for description of mixed *f-d* compounds, i.e., rare-earth ferroborates [23,24] and francisites [25]. This approach is based on the crystal field model for the rare-earth ion within mean-field approximation. In accordance with the magnetic structure and hierarchy of exchange interactions in TbTe₃, the effective Hamiltonians for $Tb^{3+}$ ions in the *i*-th sublattice ($i = 1,2$) under magnetic field $B$ can be written as:

$$\mathcal{H}_i = \mathcal{H}_i^{CF} - g_J \mu_B \mathbf{J}_i [\mathbf{B} + \lambda_{ffij} \mathbf{M}_i], \tag{2}$$

where $\mathcal{H}_i^{CF}$ is the crystal field Hamiltonian, $g_J$ is the Lande factor, $\mathbf{J}_i$ is the operator of angular moment of $Tb^{3+}$ ion, $\lambda_{ffij}$ ($j = 1,2$) are the molecular constants of intra-sublattice interlayer and intralayer interactions in accordance with Fig. 7.

The magnetic moments of rare-earth ions in the *i*-th sublattice ($i = 1,2$) are

$$\mathbf{M}_i = g_J \mu_B \langle \mathbf{J}_i \rangle. \tag{3}$$

The molecular constants $\lambda_{ff,j}$ in Eqn. (2) are related to the parameters $J_{i,j}$ in the Hamiltonian $H_{ff} = J_{i,j} S_{i(1,2)} S_{i(1,2)}$ by expressions $\lambda_{ff,j} = \frac{(g_J - 1)^2}{g_J^2 \mu_B^2} J_{i,j}$, obtained through the relation $\mathbf{S} = (g_J - 1)\mathbf{J}$ and expression for $\mathbf{M}_i$.

The crystal field Hamiltonian $\mathcal{H}^{CF}$ can be written in terms of irreducible tensor operators $C_q^{(k)}$ (see e.g. Eq. (3) in [24]). No crystal field parameters $B_q^k$ are known for $Tb^{3+}$ ions in TbTe₃. Moreover, there is no information on the parameters $B_q^k$ for any other rare-earth tellurides. To describe the low-temperature thermodynamic properties of rare-earth compounds we have to consider the ground multiplet only.

In order to determine the crystal field parameters $B_q^k$ (see Eq. (3) in [22]) we used the experimental data of the magnetic susceptibility $\chi(T)$ and magnetization curves $M(B)$ along the main crystallographic axes and data of torque signals $\tau(\varphi)$ at various temperatures. Initial crystal field parameters, from which we provided the minimization of the corresponding target function was started from the parameters for $Er^{3+}$ ion in an orthorhombic crystal Er₂BaNiO₅ [26] and for $Yb^{3+}$ ion in an orthorhombic crystal Cu₃Yb(SeO₃)₂O₂Cl [24]. The $Er^{3+}$ ion is close to $Tb^{3+}$ one in the rare-earth series. We used the following set that provides the best description of the experimental data ($B_q^k$, in cm⁻¹):

$$\begin{aligned} &B_0^2 = -20,\ B_2^2 = -117,\ B_0^4 = -596,\ B_2^4 = 1220,\ B_4^4 = 413, \\ &B_0^6 = 1863,\ B_2^6 = 1182,\ B_4^6 = -1992,\ B_6^6 = 1136. \end{aligned} \tag{4}$$

This approach allows calculating the magnetization, susceptibility and $\tau(\varphi)$ curves in TbTe₃. Obtained results of magnetization and torque are qualitative in the absence of infrared spectroscopy characterization. To calculate the magnetic properties of TbTe₃ we used the schemes of orientation of the magnetic moments, as shown in Fig. 7.

Calculation of magnetization and susceptibility curves with the crystal field parameters given in Eq. (4) confirms the easy *ac* - plane anisotropy. At $T = 5$ K, the torque is negative, $\tau(45°) < 0$, since $M_c > M_a, M_b$ (see Fig. S4). At T = 4.2 K, the difference between the $M_c$, $M_a$, $M_b$ curves at $B < 0.5$ T is negligible which results in 2.5 times smaller $\tau(45°)$ than that at 4.5 K, which leads to double reduction of the period of $\tau$ oscillations.

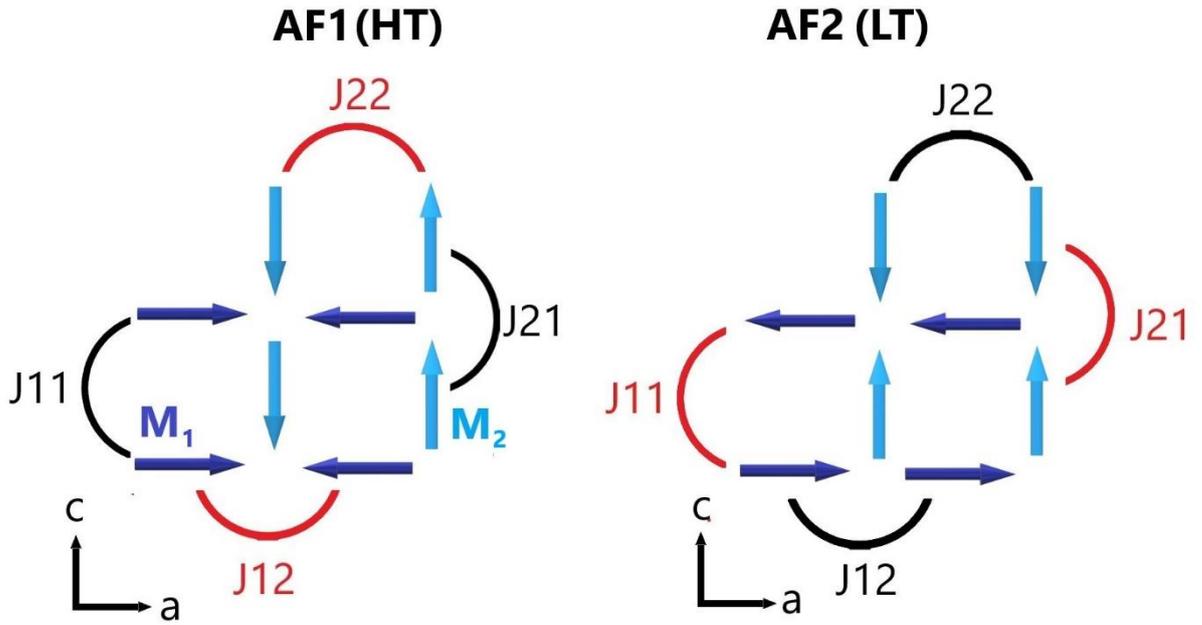

Fig. 7. The schemes of orientations of the magnetic moments of $Tb^{3+}$ ions within the bilayer used in the calculations at $B = 0$ (the axis $b$ is oriented perpendicular to the plane) in accordance with Ref. [4] in AF1 (High Temperature) and AF2 (Low Temperature) phases. $M_1$ and $M_2$ sublattices correspond to different terbium layers separated by $b/10$ along $b$ – axis within the bilayer. Red and black arcs denote ferro- and antiferromagnetic exchange interactions.

Table 1. Parameters of exchange interactions in TbTe$_3$ used for the calculation of $\tau(\varphi)$ dependencies in the ordered AF1 and AF2 phases. J denotes $J_{ff}$ in the Eq. (2). Colors correspond to that presented in Fig. 7.

| Sublattice | Phase AF1 (High Temperature) ($T$ = 4-5 K  $B$ =0.5 T) | Phase AF2 (Low Temperature) ($T$ = 2 K at $B$ =0.5 T) |
|---|---|---|
| | **FM interaction** ($J > 0$) | **AFM interaction** ($J < 0$) |
| $M_1$ | $J11_x$ = 5. K <br> $J11_y$ = 1.5 K <br> $J11_z$ = 1.8 K | $J11_x$ = -5. K <br> $J11_y$ = -3 K <br> $J11_z$ = -1. K |
| $M_2$ | $J21_x$ = 1. K <br> $J21_y$ = 1.3 K <br> $J21_z$ = 5.07 K | $J21_x$ = -1. K <br> $J21_y$ = -4. K <br> $J21_z$ = -5.22 K |
| | **AFM interaction** ($J < 0$) | **FM interaction** ($J > 0$) |
| $M_1$ | $J12_x$ = -0.5 K <br> $J12_y$ = -0.5 K <br> $J12_z$ = -0.5 K | $J12_x$ = 0.45 K <br> $J12_y$ = 0.45 K <br> $J12_z$ = 0.45 K |
| $M_2$ | $J22_x$ = -0.6 K <br> $J22_y$ = -0.6 K <br> $J22_z$ = -0.6 K | $J22_x$ = 0.51 K <br> $J22_y$ = 0.51 K <br> $J22_z$ = 0.51 K |

**Discussion**

Figure 8 demonstrates the experimental azimuthal dependence of torque at $T = 4.2$ K along with dependences typical for high temperature and low temperature regions. In the range $\varphi \approx 0 - 40°$, the $\tau(\varphi)$ is close to that in the AF phase, where $M_{a,c} < M_b$. At $\varphi \approx 40 - 90°$, where $M_{a,c} > M_b$, the $\tau(\varphi)$ curve decreases much faster than in the AF phase due to the addition of high – temperature phase possesing lower values of $\tau$ in this $\varphi$ range. All other sections of $\tau(\varphi)$ dependence can also be attributed to the dominant contributions from either PM or AF phases. In total, the appearance of two additional zero-points at 90+n° and 90-n° in $\tau(\varphi)$ dependence was attributed earlier for a reshaping of the free energy (and thus to the magnetic anisotropy) as a function of the temperature [27]. Azimuthal dependence of torque, which is proportional to the magnetization, is close to the angle dependence of charge density wave vector $q_c$ found in Ref. [6]. This gives additional confirmation of coupling of the charge density wave to the magnetic subsystem.

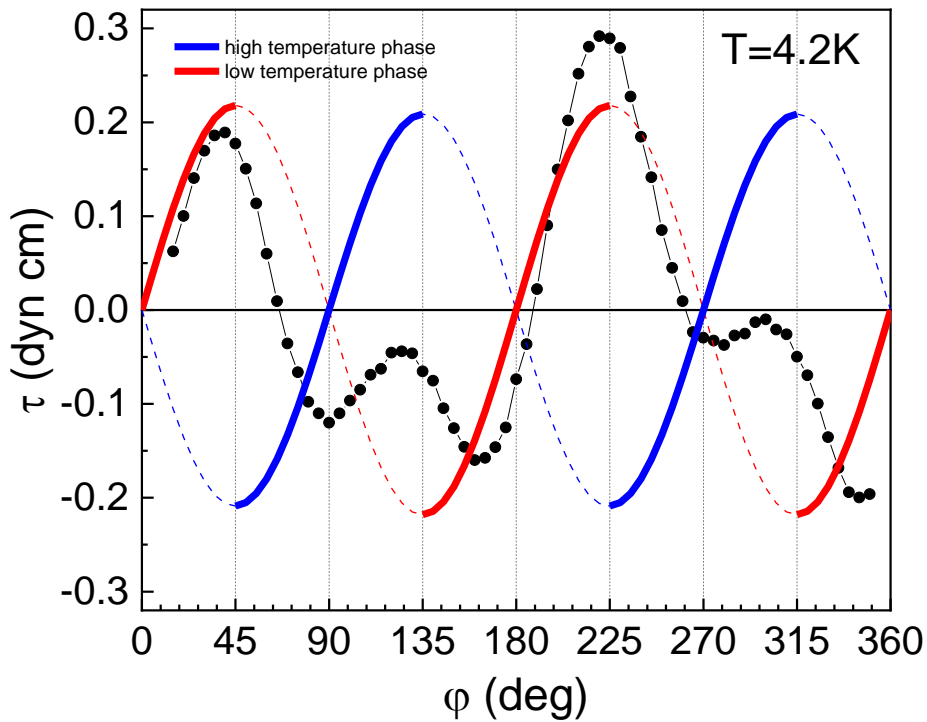

Fig. 8. The experimental azimuthal dependence of torque of TbTe$_3$ at $T = 4.2$ K (symbols). Solid and dotted lines are the simulations of $\tau(\varphi)$ at high and low temperatures.

The analysis and calculation of magnetization torque of Tb$_2$Te$_5$ is hampered by misorientation of the crystallites within the *ac*-plane. However, dominant magnetization in the *ac* plane is clearly seen in $\tau(45°) < 0$. It demonstrates two antiferromagnetic AF1 at $T_{N1} = 9.0$ K and AF2 at $T_{N2} = 6.8$ K established in thermodynamic measurements. They are higher than the corresponding 6.7 K and 5.7 K in relative TbTe$_3$ compound. This correlates with shorter Tb - Te and Te - Te distances in Tb$_2$Te$_5$. Ultrafast pump-probe spectroscopy confirms obtained magnetic phase diagram and reveals the coupling between magnetic and electron subsystems similar to TbTe$_3$.

In summary, it should be noted that terbium tellurides are the model objects to study the interaction between charge, electron and magnetic subsystems. TbTe$_3$ stands out in the row of rare earth tritellurides for three antiferromagnetic transitions of Tb subsystem and two charge density waves within Te layers along *a* and *c* - axes. Terbium ions are coordinated by nine Te anions and possess preferable orientation of magnetic moment within the *ac*-plane with non-zero component along the *b* - axis. At $T_{N1}$ = 6.7 and $T_{N2}$ = 5.7 K it experiences the formation of two commensurate antiferromagnetic structures. Below $T_{N3}$ =5.4 K, fine tuning of CDW of Te layers with AF structure induces a transition to the incommensurate phase with preferable orientation of Tb moment along the *b* – axis. Sister system, $Tb_2Te_5$, with similar $TbTe_9$ constituting polyhedra demonstrates the *ac*-plane anisotropy in the whole temperature range. It experiences two antiferromagnetic transitions. By analogy with other representatives of $RE_2Te_5$ (RE = Nd, Sm, Gd) compounds [28,29], where charge density wave vectors differ from that in the corresponding $RETe_3$ the study of charge density waves in $Tb_2Te_5$ is needed. At present, the coupling between the electron and magnetic subsystems, i.e. CDW and AF phases is obtained only in pump-probe experiment.

**Acknowledgement**

OSV thanks RSCF grant 25-12-00028 for thermodynamic measurements. The research has been carried out within the framework of the scientific program of the National Centre of Physics and Mathematics under the project "Research in strong and superstrong magnetic fields". This work has been supported by the Ministry of Science and Higher Education of the Russia within the framework of the Priority-2030 strategic academic leadership program at NUST MISIS. $Tb_2Te_5$ crystal growth has been funded by the Ministry of Science and Higher Education of the Russian Federation and Ural Federal University Program of Development within the Priority-2030 and IEM RAS FMUF-2022-0002. The study of single crystal XRD was conducted under the state assignment of Lomonosov Moscow State University, project № 121031300090-2, "Molecular and supramolecular organization of compounds, hybrid and functional materials"

**Data availability statement**

Data is available from the authors upon reasonable request.

# SUPPLEMENTARY MATERIAL

Table S1. Experimental details

| Crystal data | |
|---|---|
| Chemical formula | $Tb_2Te_5$ |
| $M_r$ | 955.84 |
| Crystal system, space group | Orthorhombic, *Cmcm* |
| Temperature (K) | 100 |
| $a$, $b$, $c$ (Å) | 4.3009 (17), 43.107 (18), 4.3016 (17) |
| $V$ (Å$^3$) | 797.5 (6) |
| $Z$ | 4 |
| Radiation type | Mo $K\alpha$ |
| m (mm$^{-1}$) | 35.41 |
| Crystal size (mm) | 0.43 × 0.34 × 0.28 |
| | |
| Data collection | |
| Diffractometer | Bruker D8 Quest |
| Absorption correction | Multi-scan<br>*SADABS2016*/2 (Bruker,2016/2) was used for absorption correction. wR2(int) was 0.1533 before and 0.069 after correction. The λ/2 correction factor is Not present. |
| No. of measured, independent and observed [$I > 2\sigma(I)$] reflections | 1591, 433, 374 |
| $R_{int}$ | 0.069 |
| $(\sin\theta/\lambda)_{max}$ (Å$^{-1}$) | 0.594 |
| | |
| Refinement | |
| $R[F^2 > 2\sigma(F^2)]$, $wR(F^2)$, $S$ | 0.090, 0.209, 1.09 |
| No. of reflections | 433 |
| No. of parameters | 30 |
| | $w = 1/[\sigma^2(F_o^2) + 831.5905P]$<br>where $P = (F_o^2 + 2F_c^2)/3$ |
| $\Delta\rho_{max}$, $\Delta\rho_{min}$ (e Å$^{-3}$) | 4.84, -3.09 |

Computer programs: *SHELXT* 2018/2 (Sheldrick, 2018), *SHELXL2018*/3 (Sheldrick, 2018).

Table S2. Atomic coordinates and equivalent isotropic displacement parameters (Å$^2$)

| Atom | x/a | y/b | z/c | $U_{eq}$ |
|---|---|---|---|---|
| Tb1 | 1.000000 | 0.40110(9) | 1.250000 | 0.0227(11) |
| Tb2 | 0.500000 | 0.30634(9) | 0.750000 | 0.0255(11) |
| Te1 | 0.500000 | 0.37944(12) | 0.750000 | 0.0221(12) |
| Te2 | 0.000000 | 0.45788(13) | 0.750000 | 0.0280(14) |
| Te3 | 0.500000 | 0.45843(12) | 1.250000 | 0.0315(15) |
| Te4 | 0.000000 | 0.32659(11) | 0.250000 | 0.0248(13) |
| Te5 | 0.000000 | 0.2506(2) | 0.750000 | 0.095(4) |

Table S3. Selected interatomic distancess (Å)

| | | | |
|---|---|---|---|
| Tb1—Te1 | 3.181 (2) | Tb2—Te4$^{iii}$ | 3.1642 (19) |
| Tb1—Te1$^{i}$ | 3.182 (2) | Tb2—Te5$^{iii}$ | 3.224 (7) |
| Tb1—Te1$^{ii}$ | 3.182 (2) | Tb2—Te5 | 3.224 (7) |
| Tb1—Te1$^{iii}$ | 3.182 (2) | Tb2—Te5$^{iv}$ | 3.264 (8) |
| Tb1—Te4$^{i}$ | 3.212 (6) | Tb2—Te5$^{v}$ | 3.264 (8) |
| Tb1—Te2$^{i}$ | 3.259 (5) | Te2—Te3 | 3.0415 (9) |
| Tb1—Te2$^{iii}$ | 3.259 (5) | Te2—Te3$^{vi}$ | 3.0415 (9) |
| Tb1—Te3 | 3.276 (5) | Te2—Te3$^{vii}$ | 3.0415 (9) |
| Tb1—Te3$^{iii}$ | 3.276 (5) | Te2—Te3$^{viii}$ | 3.0415 (9) |
| Tb2—Te1 | 3.151 (6) | Te5—Te5$^{v}$ | 3.0419 (9) |
| Tb2—Te4 | 3.1642 (19) | Te5—Te5$^{ix}$ | 3.0419 (9) |
| Tb2—Te4$^{i}$ | 3.1642 (19) | Te5—Te5$^{x}$ | 3.0419 (9) |
| Tb2—Te4$^{ii}$ | 3.1642 (19) | Te5—Te5$^{iv}$ | 3.0419 (9) |

Symmetry code(s): (i) $x+1, y, z+1$; (ii) $x, y, z+1$; (iii) $x+1, y, z$; (iv) $-x+1/2, -y+1/2, -z+1$; (v) $-x+1/2, -y+1/2, -z+2$; (vi) $x-1, y, z-1$; (vii) $x-1, y, z$; (viii) $x, y, z-1$; (ix) $-x-1/2, -y+1/2, -z+1$; (x) $-x-1/2, -y+1/2, -z+2$.

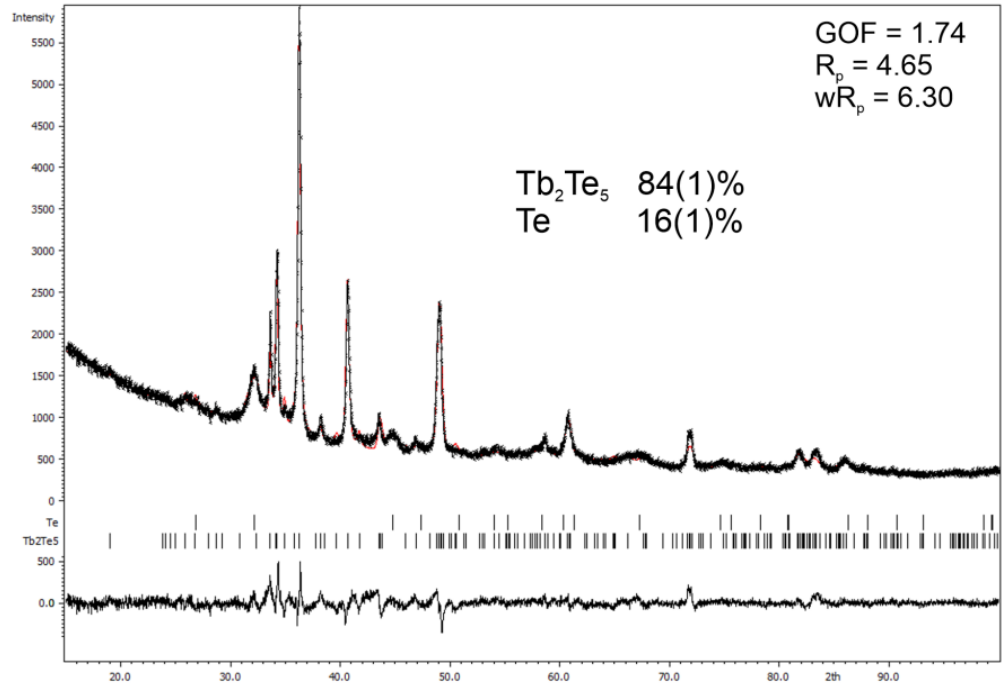

Fig. S1. The diffraction pattern of the sample obtained. The black line is the experimental pattern and the red line is the calculated pattern. The differences are shown below them. The vertical bar indicates the reflection position.

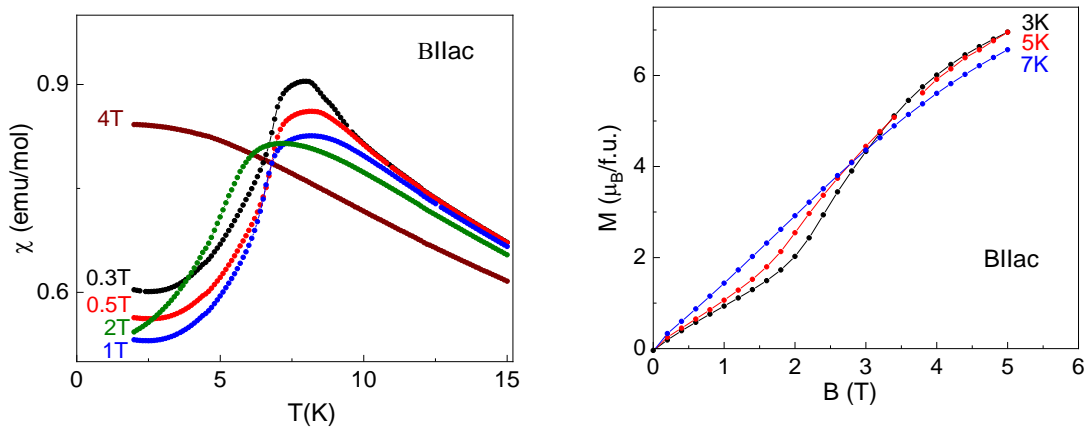

Fig. S2. Magnetization of Tb$_2$Te$_5$ crystal with magnetic field oriented within the *ac* plane versus temperature (left panel) and magnetic field (right panel).

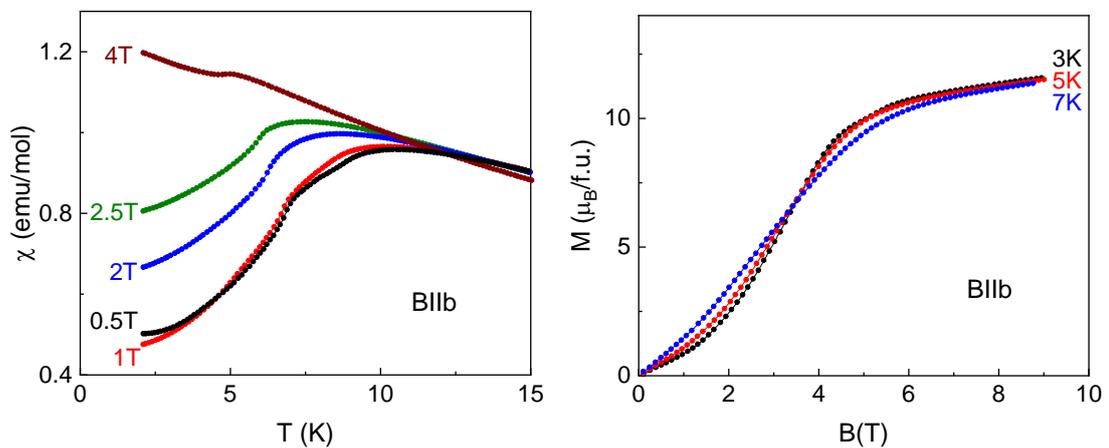

Fig. S3. Magnetization Tb$_2$Te$_5$ crystal with magnetic field orianted along the *b* - axis versus temperature (left panel) and magnetic field (right panel).

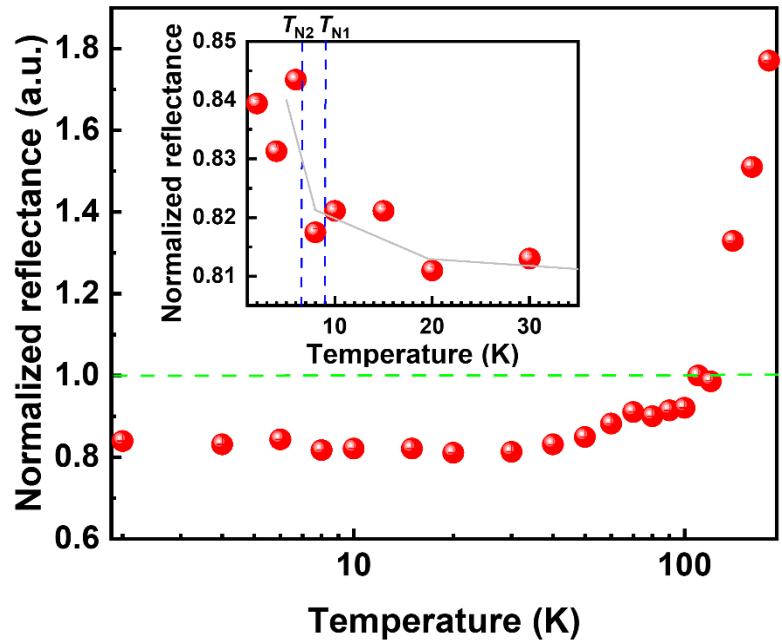

Fig. S4. Temperature-dependent static normalized reflectance (normalized to the reflectance at 110 K, where it is the zero-crossing temperature of $\Delta R/R$ in Fig. S6) of 800 nm for a Tb$_2$Te$_5$ crystal. Inset: Zoom in on the regime below 30 K, the reflectance significantly enhances as in the AFM phases ($T < T_{N1}$). The solid gray line is a guide to the eyes.

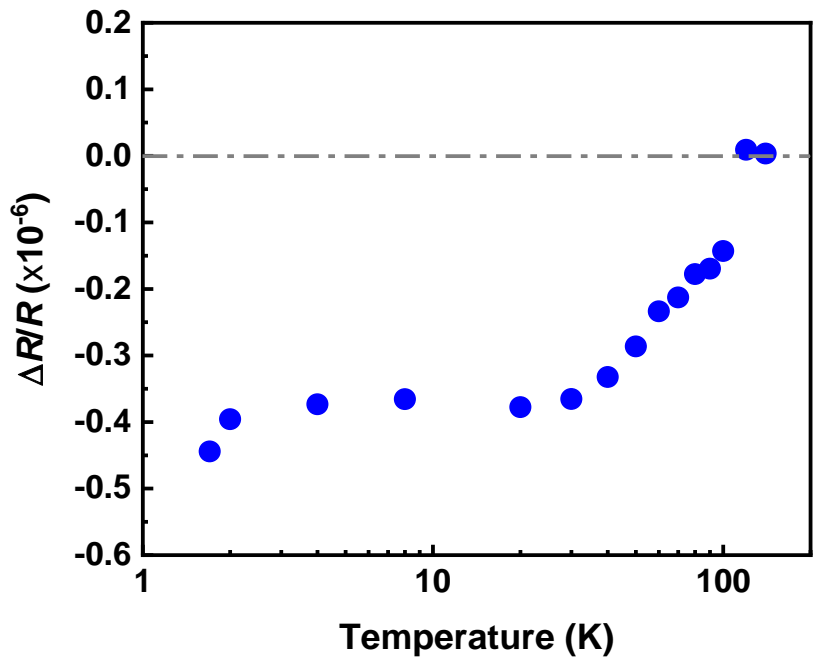

Fig. S5. The $\Delta R/R$ at 33 ps (from Fig. 4(b)) as a function of temperature.

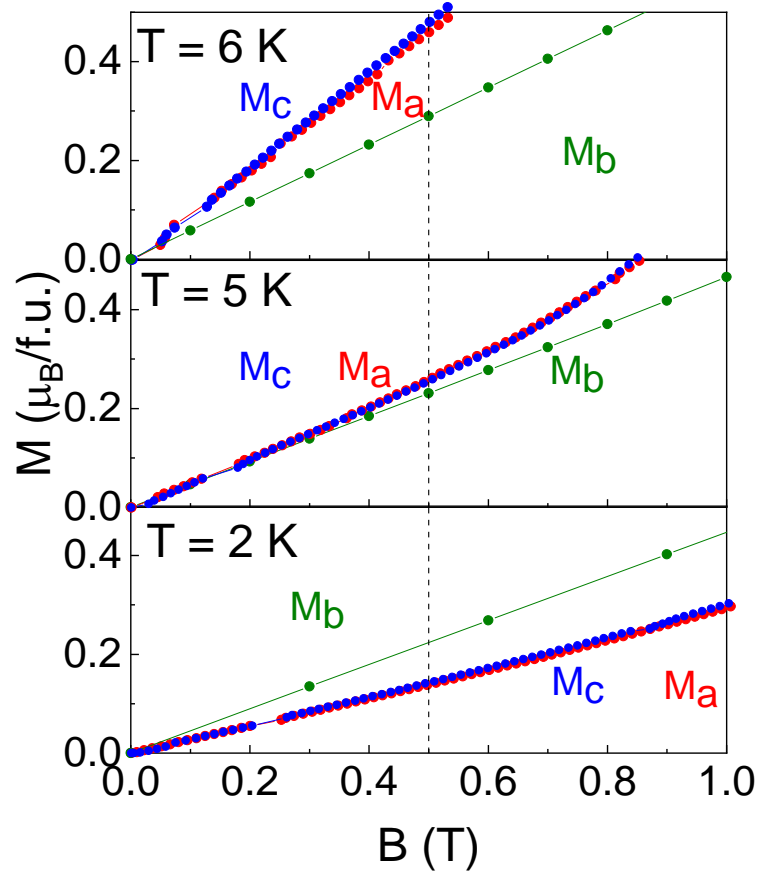

Fig. S6. Magnetization curves $M_a(B)$, $M_c(B)$ and $M_b(B)$ for TbTe$_3$ at $T = 6$, 5 and 2 K. Vertical dashed line corresponds to the field which was applied for the measurements of azimuthal dependences of torque $B = 0.5$ T in accordance with the data from Ref. [6].